\pdfoutput=1
\documentclass{PoS}

\newcommand{\code}[1]{\texttt{#1}}
\usepackage{multirow}
\usepackage{printlen}
\title{Thermal Effects on Black Hole Formation in Failed Core-Collapse
Supernovae\footnote{We acknowledge partial support by the NSF under grant
Nos. AST-0855535, OCI-0905046, and PHY-1057238.}}

\ShortTitle{Thermal Effects on Black Hole Formation in Failed
  Core-Collapse Supernovae}

\author{\speaker{Evan O'Connor}\\
  California Institute of Technology\\
       E-mail: \email{evanoc@tapir.caltech.edu}}

\author{Christian D. Ott\\
        California Institute of Technology\\
        E-mail: \email{cott@tapir.caltech.edu}}

      \abstract{We investigate several aspects of black hole formation
        in failing core-collapse supernovae using 1D
        general-relativistic hydrodynamic simulations.  We use the
        open-source code \code{GR1D} and incorporate into it
        nucleon-nucleon Bremsstrahlung, a crucial neutrino
        pair-production channel.  We focus on how various thermal
        effects can influence the postbounce supernova evolution
        towards black hole formation. By performing simulations with
        and without nucleon-nucleon Bremsstrahlung, we investigate the
        sensitivity of black hole formation to thermal support in the
        protoneutron star. We also investigate delayed black hole
        formation by artificially driving explosions in an extreme
        model where the protoneutron star is initially thermally
        supported above the maximum baryonic cold neutron star mass
        but then collapses to a black hole after neutrino cooling
        removes sufficient thermal support.}

\FullConference{11th Symposium on Nuclei in the Cosmos\\
		 19-23 July 2010 \\
		 Heidelberg, Germany.}

\begin{document}

\def\aj{{AJ\,\,}}                   
\def\araa{{ARA\&A\,\,}}             
\def\apj{{ApJ\,\,}}                 
\def\apjl{{ApJ\,\,}}                
\def\apjs{{ApJS\,\,}}               
\def\ao{{Appl.~Opt.\,\,}}           
\def\apss{{Ap\&SS\,\,}}             
\def\aap{{A\&A\,\,}}                
\def\aapr{{A\&A~Rev.\,\,}}          
\def\aaps{{A\&AS\,\,}}              
\def\azh{{AZh\,\,}}                 
\def\baas{{BAAS\,\,}}               
\def\cqg{{Class. \& Quan. Grav.\,\,}}                 
\def\jrasc{{JRASC\,\,}}             
\def\memras{{MmRAS\,\,}}            
\def\mnras{{MNRAS\,\,}}             
\def\pra{{Phys.~Rev.~A\,\,}}        
\def\prb{{Phys.~Rev.~B\,\,}}        
\def\prc{{Phys.~Rev.~C\,\,}}        
\def\prd{{Phys.~Rev.~D\,\,}}        
\def\pre{{Phys.~Rev.~E\,\,}}        
\def\prl{{Phys.~Rev.~Lett.\,\,}}    
\def\pasp{{PASP\,\,}}               
\def\pasj{{PASJ\,\,}}               
\def\qjras{{QJRAS\,\,}}             
\def\skytel{{S\&T\,\,}}             
\def\solphys{{Sol.~Phys.\,\,}}      
\def\sovast{{Soviet~Ast.\,\,}}      
\def\ssr{{Space~Sci.~Rev.\,\,}}     
\def\zap{{ZAp\,\,}}                 
\def\nat{{Nature\,\,}}              
\def\iaucirc{{IAU~Circ.\,\,}}       
\def\aplett{{Astrophys.~Lett.\,\,}} 
\def\apspr{{Astrophys.~Space~Phys.~Res.\,\,}}                               
\def\bain{{Bull.~Astron.~Inst.~Netherlands\,\,}}                            
\def\fcp{{Fund.~Cosmic~Phys.\,\,}}  
\def\gca{{Geochim.~Cosmochim.~Acta\,\,}}   
\def\grl{{Geophys.~Res.~Lett.\,\,}} 
\def\jcp{{J.~Chem.~Phys.\,\,}}      
\def\jgr{{J.~Geophys.~Res.\,\,}}    
\def\jqsrt{{J.~Quant.~Spec.~Radiat.~Transf.\,\,}}                           
\def\memsai{{Mem.~Soc.~Astron.~Italiana\,\,}}                               
\def\nphysa{{Nucl.~Phys.~A\,\,}}   
\def\physrep{{Phys.~Rep.\,\,}}   
\def\physscr{{Phys.~Scr\,\,}}   
\def\planss{{Planet.~Space~Sci.\,\,}}   
\def\procspie{{Proc.~SPIE\,\,}}   

\section{Introduction}

The majority of the stellar-mass black holes (BHs) present in the
modern universe are expected to form by the collapse of protoneutron
stars in core-collapse supernovae (CCSNe).  There are several channels
through which this can occur: {\emph{(i)}} the supernova shock fails
to be reenergized and accretion pushes the PNS over its maximum mass,
leading to collapse (c.f. \cite{sumiyoshi:07, sumiyoshi:09,
  fischer:09a, oconnor:10b}), {\emph{(ii)}} fallback after a
successful CCSN can lead to PNS collapse \cite{zhang:08}, and
{\emph{(iii)}} nuclear phase transitions or PNS cooling may cause the
PNS to collapse during its cooling phase (e.g., \cite{baumgarte:96b}).

In this contribution to the proceedings of the 11th Symposium on
Nuclei in the Cosmos, we investigate several aspects of BH formation
in CCSNe that are related to thermal effects.  In \S\ref{sec:methods},
we briefly outline the simulation code we use to study core collapse
and BH formation and discuss our implementation of nucleon-nucleon
Bremsstrahlung, a recent addition to our code. In
\S\ref{sec:resultsNNB}, we show that in failing CCSNe this interaction
increases the neutrino luminosity, removes thermal support, and causes
the PNS to collapse to a BH earlier than when the interaction is not
included. Finally, in \S\ref{sec:resultsdelayedBH}, we artificially
drive explosions in which the resultant PNS is thermally supported
above the maximum baryonic PNS mass.  Using this approach with an
extreme model, we demonstrate delayed BH formation due to PNS cooling.

\section{Methods}
\label{sec:methods}
\vspace*{-0.25cm}
\subsection{\code{GR1D}}
We use the open-source, spherically-symmetric, general-relativistic
(GR) Eulerian hydrodynamics code \code{GR1D}, which is available at
\mbox{\tt http://stellarcollapse.org}. We refer the reader to
Ref.~\cite{oconnor:10} for the full details of \code{GR1D} and give
here only a brief description.  \code{GR1D} follows the formalism of
Romero~et~al. \cite{romero:96}, which uses the radial gauge and polar
slicing gauge condition in spherically-symmetric GR. This reduces the
metric to a diagonal form with two metric functions that are
determined by hydrodynamic variables.  \code{GR1D} is designed to
follow the evolution of stars beginning from the onset of core
collapse to BH formation.  Its GR hydrodynamics module utilizes high
resolution shock capturing methods. \code{GR1D} makes use of several
microphysical equations of state (EOS), also available online at
\mbox{\tt http://stellarcollapse.org}, and uses a 3-neutrino
leakage/heating scheme (described in \S\ref{sec:leakageintro}).  Here,
we make use of the EOS from Lattimer \& Swesty (1991) \cite{lseos:91}
with an incompressibility of $K_0=220\,$MeV (LS220).

\subsection{Neutrino Treatment \& Nucleon-Nucleon Bremsstrahlung}
\label{sec:leakageintro}
\code{GR1D} uses a hybrid neutrino leakage scheme based on the schemes
of \cite{ruffert:96} and \cite{rosswog:03b} and includes the standard
reactions therein.  It includes three neutrino types, $\nu_e$,
$\bar{\nu}_e$, and $\nu_x = \{\nu_\mu, \bar{\nu}_\mu, \nu_\tau,
\bar{\nu}_\tau\}$.  The neutrino emission rates, both energy and
number, are determined from an interpolation between the diffusion and
free streaming limits.  These emission rates are combined with a
parameterized neutrino heating scheme to calculate the total energy
emitted from a zone. The parameterization of our heating scheme, via
the scale factor $f_\mathrm{heat}$, allows us to drive artificial
explosions.  Full details of \code{GR1D}'s neutrino leakage/heating
scheme can be found in \cite{oconnor:10}. An application of our
parameterized heating scheme can be found in \cite{oconnor:10b}.

A process for neutrino pair production not included in theleakage
schemes of \cite{ruffert:96,rosswog:03b} is nucleon-nucleon
Bremsstrahlung (NNBrems), i.e. $N + N \to N + N +
\nu\bar{\nu}$. NNBrems dominates the neutrino pair production at high
matter densities.  In CCSNe, this regime is very optically thick and
NNBrems is a subdominant contributor to the total neutrino luminostiy.
However, during PNS cooling or NS-NS or NS-BH mergers, emission from
NNBrems may be considerable.  As a thermal neutrino pair-production
process, we include NNBrems into \code{GR1D} to investigate its effect
on BH formation.  We use the single neutrino pair emissivity for
NNBrems from \cite{brt:06}\footnote{\cite{brt:06} mistakenly included
  the $\zeta$ factor in the numerical constant as well as in the
  expression for $Q_{nb}$. We use the corrected expression here
  (Burrows, {\emph{private communication}}, 2010).},
\begin{equation}
Q_{nb} = 2.0778\times10^{30}\zeta \left(X_n^2+X_p^2+{28 \over
  3}X_nX_p\right)\rho_{14}^2\left({T \over
  {\rm{MeV}}}\right)^{5.5}\ {\rm{ergs}}\ {\rm{cm}}^{-3}\ {\rm{s}}^{-1}\,,\label{eq:NNBrems}
\end{equation}  
where $\rho_{14}$ is the matter density in units of
10$^{14}\,$g~cm$^{-3}$, $T$ is the matter temperature in MeV, $X_n$
and $X_p$ are the mass fractions of neutrons and protons respectively
and $\zeta$ is a factor that absorbs all interaction ambiguities. We
adopt the value of $\zeta=0.5$ following \cite{brt:06}. For
\code{GR1D}, we require the single-neutrino emissivities (both energy
and number). Absorptive opacities are not needed (but could be
obtained via Kirchhoff's law), since \code{GR1D}'s leakage scheme
includes opacity contributions only from scattering and from
charged-current absorption processes.  With the publication of this
proceedings article, we have updated \code{GR1D} (to v1.03) to include
NNBrems as an option in the leakage scheme.  

\section{Results}
\vspace*{-0.25cm}
\subsection{Nucleon-Nucleon Bremsstrahlung}
\label{sec:resultsNNB}
We perform 12 simulations in which we investigate the effect of
NNBrems in BH-forming core collapse simulations.  We choose 6
progenitor models from the ``$u$ series'' of Woosley~et~al. (2002)
\cite{whw:02}, namely the 35, 40, 45, 50, 60, and 75$\,M_\odot$ models
with metallicities of $10^{-4}\,Z_\odot$. These models have large iron
cores at the onset of core collapse and, as we shall see, form BHs
very soon after bounce.  They yield large temperatures in their PNSs
and therefore are ideal for studying thermal effects.  For each model
we run simulations with and without the NNBrems emissivities
(Eq.~\ref{eq:NNBrems}) included in our neutrino leakage scheme.
Table~\ref{tab:NNBrems} presents the results of these simulations, we
include the time to BH formation and the total neutrino energy emitted
between bounce and BH formation.

\begin{table}[h]
\centering
\begin{tabular}{cc|cccccc}
  \hline
  \hline
  \multicolumn{2}{c|}{\bf Model} & {\bf $u$35} & {\bf $u$40} & {\bf $u$45} & {\bf $u$50} & {\bf $u$60} &
  {\bf $u$75} \\
  \hline
  \multirow{2}{*}{NNBrems} & $t_\mathrm{BH}$ [s] & 0.522 & 0.478 & 0.578
  & 0.747 & 0.630 & 0.311 \\ 
  & $\nu$-energy [B] & 257 & 260 & 277 & 281 & 233 & 184 \\ 
  \hline
  \multirow{2}{*}{No NNBrems} & $t_\mathrm{BH}$ [s] & 0.526 & 0.482 &
  0.582 & 0.749 & 0.635 & 0.311 \\
  & $\nu$-energy [B] & 249 & 253 & 268 & 273 & 229 & 179 \\
  \hline
\end{tabular}
\caption{Effect of NNBrems on time to BH formation and total emitted
  neutrino luminosity until BH formation. The progenitor models are taken from the
  $10^{-4}\,Z_\odot$ model set of \cite{whw:02} and simulated with the
  LS220 EOS.  $1\,$B = $10^{51}\,$ergs.}\label{tab:NNBrems}
\end{table}

NNBrems results in a slightly higher\footnote{NNBrem  is a
  more dominant contributor to the neutrino luminosity in lower ZAMS
  mass progenitors, i.e. $8\,M_\odot \lesssim M \lesssim 20\,M_\odot$.
  In these models the temperatures are typically lower and NNBrem,
  which scales  $\propto T^{5.5}$, becomes comparable to other neutrino
  pair-production processes, which scale  $\propto T^9$.} ($\sim$3\%)
average neutrino luminosity as is demonstrated by the higher total
energy emitted in neutrinos.  This is due to the increased neutrino
emissivity in the dense core of the PNS where NNBrems dominates. The
extra loss of thermal energy from this region reduces the thermal
pressure, thus removes thermal pressure support from the PNS.  This,
in turn, lowers the maximum PNS mass and causes the dynamical collapse
of the PNS to begin at somewhat earlier times ($\Delta t \sim\,${{\it
    O}}(ms)) compared to the case with no NNBrems.

\subsection{Thermal Effects}
\label{sec:resultsdelayedBH}
Due to thermal pressure support, the maximum gravitational mass of a
PNS in a CCSN environment can be significantly higher than the cold
neutron star (CNS) value \cite{oconnor:10b}. In extreme models with
soft EOS, or in models with a hyperonic EOS, it is also possible that
the {\emph{baryonic}} mass of the PNS can exceed its CNS value.  An
interesting case arises when, during a core-collapse event, the shock
is successfully reenergized but enough material has accreted onto the
PNS so that it exceeds its maximum CNS baryonic mass.  Initially
stable, the PNS cools via neutrinos. Since baryonic mass is conserved,
once sufficient thermal support is lost, no stable configuration
exists and the PNS will collapse to a BH.
\begin{figure}[t]
\begin{center}
  \includegraphics[width=210pt]{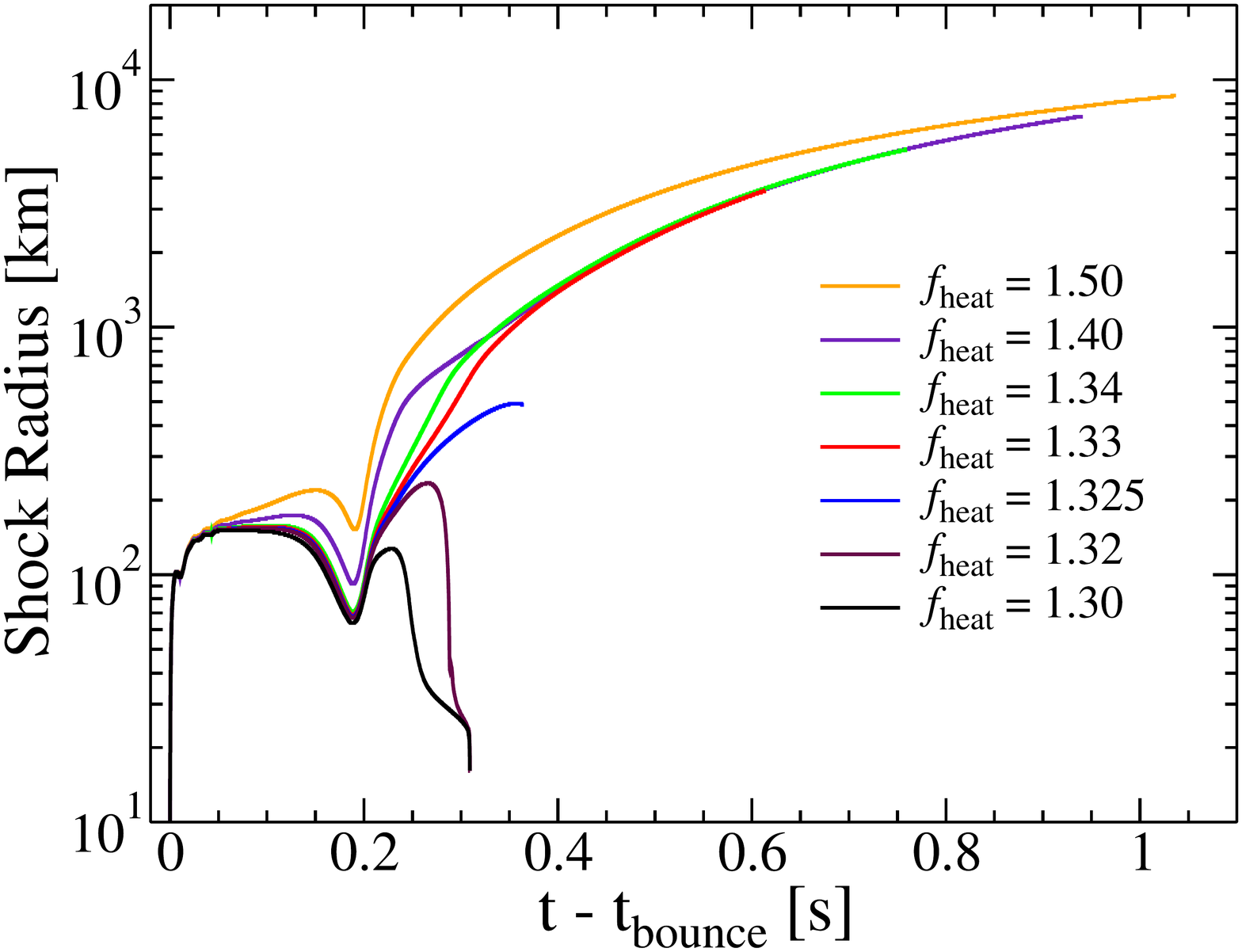}
\includegraphics[width=210pt]{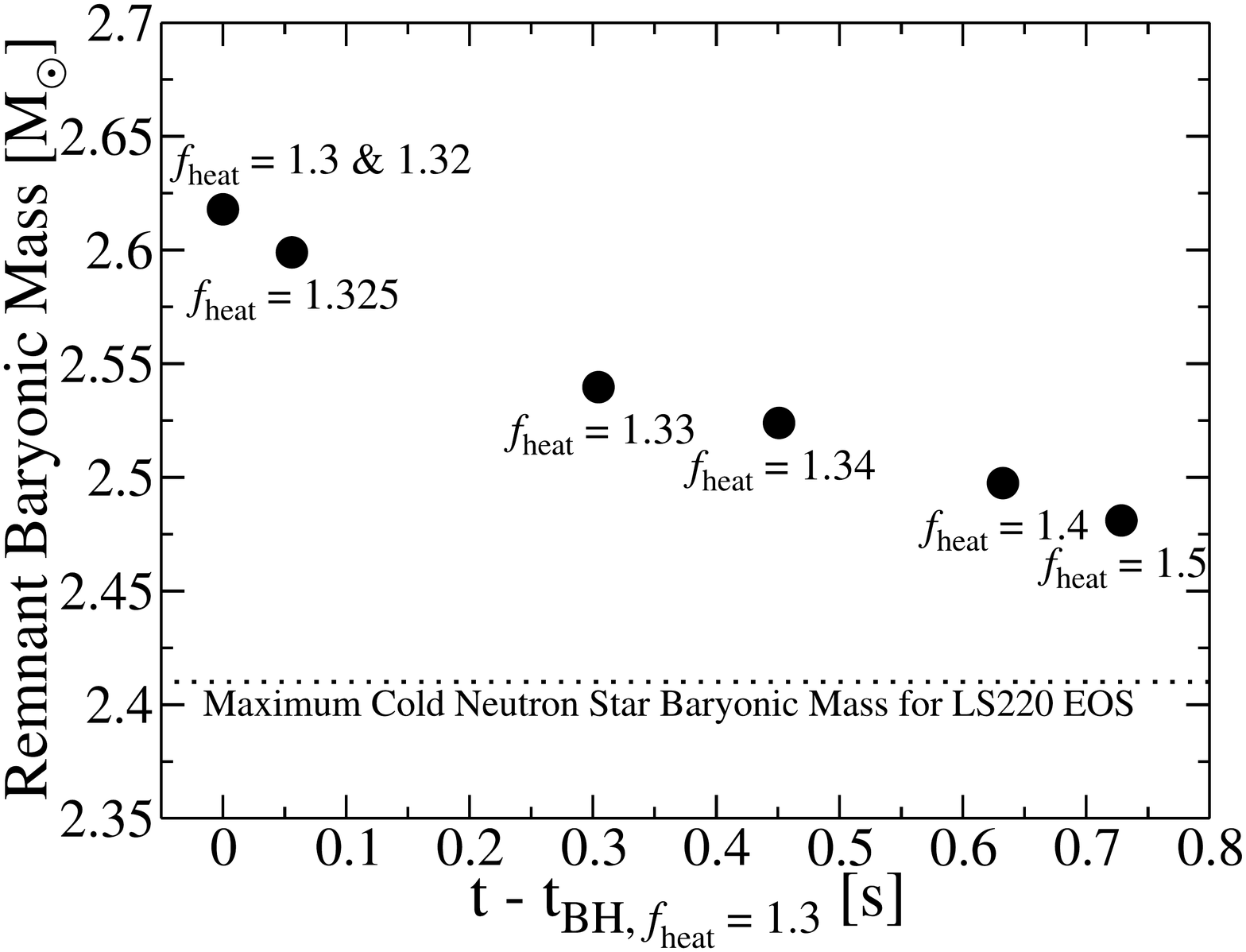} 
\caption{Shock radii ({\emph{left panel}}) for the $u$75 model of
  \cite{whw:02} with 7 values of $f_\mathrm{heat}$.  For each value of
  $f_\mathrm{heat}$, we plot in the {\emph{right panel}} the baryonic
  mass of the PNS after accretion has ceased versus the time each
  model remains stable. Less massive PNS remain stable for longer
  times.  The dotted line indicates the maximum cold neutron star
  baryonic mass for the LS220 EOS.}\label{fig:delayedmass}
\end{center}
\vspace*{-0.5cm}
\end{figure}
We perform simulations with the $u$75 model from \cite{whw:02} and we
artificially adjust the amount of neutrino heating via the scale
factor $f_\mathrm{heat}$ (see \cite{oconnor:10} for details) to
achieve explosions. The $u$75 is an extreme model. In a failing CCSN,
it forms a BH roughly $\sim$300~ms after bounce with a baryonic
(gravitational) mass of $\sim$2.62 ($\sim$2.44)$\,M_\odot$.  For
comparison, the LS220 EOS has a maximum CNS baryonic (gravitational)
mass of $\sim$2.41 ($\sim$2.04)$\,M_\odot$. Due to the presence of a
compositional interface located at a baryonic mass coordinate of $\sim
2.5\,M_\odot$, where the density drops by $\sim$50\%, it is possible
to launch explosions very close to the time of PNS collapse and
achieve the scenario discussed above.  In the right panel of
Fig.~\ref{fig:delayedmass}, we show the baryonic mass of the PNS at BH
formation for 7 different values of $f_\mathrm{heat}$. The left panel
shows the shock position until BH formation for these same models. In
the $f_\mathrm{heat} = 1.30$, 1.32, and 1.325 models, the PNS never
ceases to accrete material.  In the remaining models, as
$f_\mathrm{heat}$ increases, the shock is reenergized at earlier times
leaving PNSs with increasingly smaller baryonic masses.  As neutrinos
extract thermal energy from the PNS, the latter becomes unstable and
collapses to a BH.  The timescale for neutrino cooling is typically
10-100~s, but in the case of the extreme $u$75 model, BH formation
happens within $\sim 1\,$s for these values of $f_\mathrm{heat}$. For
this particular model, we find that NNBrems does not have an
observable effect due to the extremely high temperatures
($50-100\,$MeV) in the outer regions of the PNS.  As mentioned in
\S\ref{sec:resultsNNB}, very high temperatures lead to the dominance
of other neutrino pair-production processes over NNBrems at early
post-explosion times.  As an aside, hyperonic EOS can have a much
lower maximum CNS baryonic mass but still support large PNS baryonic
masses when the temperature is high \cite{sumiyoshi:09}.  In this
case, the above scenario can occur for a much larger range of
progenitors and extreme models do not need to be invoked.

{\tiny

}

\end{document}